\documentclass[a4paper]{article}
\usepackage{graphicx}
\usepackage{times}
\usepackage{amssymb}
\usepackage[sort,square]{natbib}
\usepackage[vcentering,dvips]{geometry}
\def\gsim { \lower .75ex \hbox{$\sim$} \llap{\raise .27ex \hbox{$>$}} }
\def\lsim { \lower .75ex \hbox{$\sim$} \llap{\raise .27ex \hbox{$<$}} }

\begin{document}

\title{The Effect of Tidal Stripping on Composite Stellar Populations in Dwarf Spheroidal Galaxies}

\author{
{Laura V. Sales$^{1}$,
Amina Helmi$^{1}$
and Giuseppina Battaglia$^{2}$
}
\\
\\
$^{1}$ Kapteyn Astronomical Institute, P.O. Box 800, Groningen, The Netherlands\\
$^{2}$European Organisation for Astronomical Research in the Southern Hemisphere, \\
Karl-Schwarzschild-Strasse 2, 85740 Garching bei Muenchen, Germany\\
}
\maketitle
\begin{abstract}
We use N-body simulations to study the effects of tides on the
  kinematical structure of satellite galaxies orbiting a Milky
  Way-like potential. Our work is motivated by observations of dwarf
  spheroidal galaxies in the Local Group, for which often a
  distinction is possible between a cold centrally concentrated metal
  rich and a hot, extended metal poor population.  Here we focus on
  the evolution of a spherical satellite with two stellar components
  set {\it ab-initio} to be spatially and kinematically segregated,
  and which are embedded in an extended dark matter halo.  We find
  that an important attenuation of the initial differences in the
  distribution of the two stellar components occurs for orbits with
  small pericentric radii ($r_{per} \le 20$ kpc).  This is mainly due
  to: $i)$ the loss of the gravitational support provided by the dark
  matter component after tidal stripping takes place, which forces a
  re-configuration of the luminous components, and $ii)$ tides
  preferentially affect the more extended stellar component, leading
  to a net decrease in its velocity dispersion as a response for the
  mass loss, which thus shrinks the kinematical gap.  We apply these
  ideas to the Sculptor and Carina dwarf spheroidals. Differences in
  their orbits might help to explain, under the assumption of similar
  initial configurations, why in the former a clear kinematical
  separation between metal poor and metal rich stars is apparent,
  while in Carina this segregation is significantly more subtle.
\end{abstract}

\section{Introduction}
Dwarf galaxies are by number the most common kind of galaxies in the
Local Group. Among all types of dwarf
galaxies, such as dwarf irregulars, dwarf ellipticals, transition
types and dwarf spheroidal (dSphs), the latter dominate the satellite 
population of the large spirals.

Dwarf spheroidal galaxies are faint (with luminosities between
$10^3-10^7 L_\odot$), metal poor and tend not to be rotationally
supported. The dark matter mass enclosed within the optical extent of
these small objects ranges between $10^5-10^8 M_\odot$
\citep{walker07}, which extrapolated at large radii would give virial
masses of the order of $10^8-10^9 M_\odot$ and higher
\citep{strigari08}. The star formation histories of these small
objects are known to be complex and vary from object to object, but
all dSphs contain ancient stellar populations, with ages $>$10 Gyr
old.  In most dSphs the ancient stellar component is the dominant one
or even the only one present (e.g. Sculptor, Draco, Ursa Minor); there
are however a few cases where intermediate age stars (3-6 Gyr old) are
also present (such as Carina) or dominate the overall stellar
population (e.g. Fornax, Leo~I) \citep{gallagher94,mateo98,grebel03}.

An intriguing observational feature of dSphs is that, notwithstanding
their small sizes, the star formation and chemical enrichment
histories are not uniform inside these objects, but varied as
exemplified by the presence of spatial gradients in the average
metallicity and sometimes age properties of their stellar populations.
Harbeck et al. (2001) \citep{harbeck01} used wide-field imaging to
study a set of 9 Local Group dwarfs and examined the spatial
distribution of stars in different evolutionary phases as selected
from the colour-magnitude diagram (CMD).  In particular, the authors
used horizontal branch (HB) stars, known to be ancient ($>10$ Gyr
old), and determined the relative spatial distribution of the blue and
red (horizontal branch) populations (BHB and RHB, respectively) in
order to identify the existence of metallicity/age variations. Harbeck
et al. found that segregation of populations may not be uncommon in
dSphs although it is not {\it necessarily} present in all of them.

Further new evidence on the presence of multiple populations came a
couple of years later, when Tolstoy et al. (2004) \citep{tolstoy04}
added to wide-field imaging also the information from wide-field
spectroscopy, which yielded metallicities and line-of-sight velocities
for hundreds individual stars in the Sculptor dSph.  These authors
found that metal rich stars in Sculptor are more centrally
concentrated and have on average lower velocity dispersions than those
of the metal poor component, which on the contrary, show hotter
kinematics and are more extended. Clear stellar population gradients
have been confirmed for Tucana, Sextans, Sculptor \& AndVI
\citep{harbeck01}, and more recently in Draco and AndII
\citep{faria07,mc07b}. On the other hand, weak or no gradients have
been found in other satellite galaxies of the Local Group as well,
such is the case for Carina, Leo I \citep{koch07}, AndIII \& AndV$-$VI
\citep{harbeck01}. Carina for example shows a mild segregation between
HB and red clump stars, comparable in amplitude to the metallicity
gradient found for spectroscopically confirmed Carina red giant stars
\citep{koch06}.  When kinematical information is also available it is
found that these stellar populations differences sometimes clearly
show up as multiple kinematical components as in the case of Sculptor
and Fornax (ref. \citep{battaglia06}). In other cases such as in
Carina the kinematical distinction of stars with different
metallicities is more subtle.

The eventual occurrence and evolution of this spatial and kinematical
segregation of stars in dwarf galaxies constitutes an interesting
problem for models of galaxy formation.  Tolstoy et al (2004) put
forward several hypotheses in order to explain their findings in
Sculptor: two episodes of star formation with an inactive period
between them, the accretion of gas from the dwarf surroundings able to
cool and condensate in the core of them or the photo-evaporation of the
outer gas layers due to a UV background \citep{tolstoy04}.  In recent
work, Kawata et al. (2006) \citep{kawata06} followed the formation of
a dwarf galaxy in a self$-$consistent cosmological numerical
simulation up to redshift $\sim 6$. These authors found that the
complex interplay between star formation and feedback, coupled to
chemical evolution, resulted in a system with a marked metallicity
gradient. Kawata et al. thus proposed that a sufficiently steep
continuous gradient of a {\it single} stellar population may appear to
observations as two kinematically different metal-poor and metal-rich
populations.  Even though this constitutes an interesting possibility,
the metallicity distribution and velocity dispersion profiles in this
model are only marginally consistent with observations, and the system
may evolve further (for example in mass) from redshift 6 to the
present-day.

Tides might also play an important role on the evolution of multiple
stellar components in dSphs. Given the proximity to the primary
galaxy, tidal effects are likely to be relevant on satellite galaxies
such as the MW dSph. In some models, the strong tidal field induced by
the proximity to the primary may lead to fundamental changes in the
dwarfs internal configuration. These include the development of bars
and bending instabilities that strongly affect the internal kinematics
of the satellite, transforming rotationally supported systems into hot
spheroidals \citep{mayer01a,mayer01b}.  Such processes are triggered
especially for eccentric orbits, implying that the properties of the
dwarfs we observe today around bright galaxies could possibly give us
indirect clues about their primordial (original) state, i.e. before
interactions with the primary shaped them differently.

In this paper we explore how gravitational effects may induce a mixing
of two initially distinct stellar components in dwarf spheroidal
galaxies orbiting a MW-like host potential. Our aim is to gauge the
evolution of multiple component dwarfs as they move through the host
potential and how this depends on their specific orbital path.  In
particular, we focus on the differences shown when the same fiducial
satellite is placed and evolved on four orbits around the host with
different periods and pericenter distances. We therefore do not
attempt to explain the origin of such metallicity/kinematical
segregation in the dwarfs in the present paper.

We describe the numerical experiments in Section \S~\ref{sec:models},
present and discuss our results in Section \S~\ref{sec:results} and
summarise our findings in Section \S~\ref{sec:concl}.


\section{Numerical experiments}
\label{sec:models}

Observations of the radial velocities of stars in dSph suggest that
these are in general dark matter dominated systems, with only a very
small ($\leq 5\%$) fraction of their baryonic mass in gaseous form
\citep{gallagher94,mateo98,grebel03}.  This allows a reasonable
modeling of their present-day properties and recent evolution by means
of collisionless N-body simulations. Here we use GADGET$-$2
\citep{springel05} to simulate satellite galaxies orbiting in a
(static) Milky Way-like halo. Specifically, we focus on the evolution
of satellites with two spherical stellar components embedded in an
extended dark matter halo.

\subsection{The model for the host potential}
We model the (Milky Way) host potential as a (fixed) three component system with :

\begin{itemize}
\item a Navarro, Frenk and White \citep{nfw97} dark matter halo:

$$\rho(r) \propto \frac{1}{x(1+cx)^2}$$

with mass $M_{dm}=1 \times 10^{12} \rm M_\odot$ and concentration c=12 (Klypin et al. 2002) \citep[ref ][]{klypin02},

\item a Hernquist \citep{hernquist90} bulge:

$$\rho(r) = \frac{M_{blg}d}{2\pi r}\frac{1}{(r+d)^3} $$

with mass and scale length: $M_{blg}=3.4 \times 10^{10} M_\odot$ 
and $d=0.7$ kpc, 

\item a Miyamoto$-$Nagai \citep{miyamoto_nagai} disk:

$$\rho(R,z)=\frac{b^2M_{dsk}}{2\pi} \frac{aR^2+(a+3\sqrt{z^2+b^2})(a+\sqrt{z^2+b^2})^2}{[R^2+(a+\sqrt{z^2+b^2})^2]^{5/2}(z^2+b^2)^{3/2}}$$

with parameters: $M_{dsk}=1 \times 10^{11} M_\odot$, $a=6.5$ kpc and
$b=0.26$ kpc (Johnston et al. 1999) \citep{johnston99}.
\end{itemize}

\subsection{The model for the satellite}

The satellite model consists of two baryonic components following
Plummer density profiles:

$$\rho_i = \frac{3M_i b_i^2}{4\pi} \frac{1}{(r^2 + b_i^2)^{5/2}}$$ 
of different masses $M_i$ and scale lengths $b_i$ that mimic a
concentrated metal rich and more extended metal poor stellar
components ($C_1$ and $C_2$, respectively). Both spheres are embedded
within a Hernquist-profile dark matter halo.  The masses of each of
these components are set to $1.5, 3.5 \times 10^6$ and $2 \times 10^9
\rm M_\odot$ for $C_1$, $C_2$ and dark halo. The Plummer spheres scale
lengths for $C_1$ and $C_2$ are $b_1=0.11 \rm kpc$ and $b_2=0.35 \rm
kpc$, respectively. For the dark matter, the scale radius is $a = 2.7
\rm kpc$, which gives a half mass radius $\sim 6.5 \rm kpc$.

The satellite galaxy is modelled using $\sim 310000$ particles
distributed in a dark matter halo (200000) and two luminous components
(110000) set by construction to have an initial spatial and kinematic
segregation.  The procedure to generate the initial condition for
simulations of compound galaxies was developed by Hernquist (1993) 
\citep{hernquist93} and invokes the moments of the Collisional
Boltzman Equations (CBE) to self-consistently approximate the (unknown) distribution
functions of particles in all galaxy components. We
briefly describe this procedure in what follows and refer the reader
to the original paper (ref \citep{hernquist93}) for further details.

For an isotropic, spherically symmetric system, the second moment
 of the CBE can be written as:
$$ \langle v_r^2\rangle = \frac{1}{\rho} \int^\infty_r \rho(r) \frac{d\Phi}{dr} dr$$
where $\langle v_r^2 \rangle$ is the velocity dispersion in the radial
direction, $\rho$ is the mass density distribution and $\Phi$ the
total gravitational potential of the system. This formula can be
extended to the case where several subcomponents coexist in mutual
equilibrium.  For such configurations, the second velocity moment of
the component $C_j$ would be given by:

$$ \langle v_{r,j}^2 \rangle = \frac{1}{\rho_j} \int^\infty_r \sum_{i=1}^{n_c} \rho_j \frac{d\Phi_i}{dr} dr$$
(where $j$ and $i$ vary over the $n_c$ different components of the
system).  Once the density profiles $\rho_j$ for each of the
components have been fixed, the solution to this equation may be
found, either analytic or numerically. To set up the particles
velocities we assume isotropic Gaussian distribution functions.  This
procedure then allows us to assign a velocity $v$ to a particle
located at a distance $r$ in component $C_j$, where $v$ is taken
randomly from the distribution:

$$F_j(r,v) = 4\pi (\frac{1}{2\pi\sigma_j^2})^{3/2} v^2 exp(-v^2/2\sigma_j^2)$$
with dispersion $\sigma_j =  \langle v_{r,j}^2 \rangle$.

\begin{figure*}
\begin{center}
\includegraphics[width=0.485\linewidth]{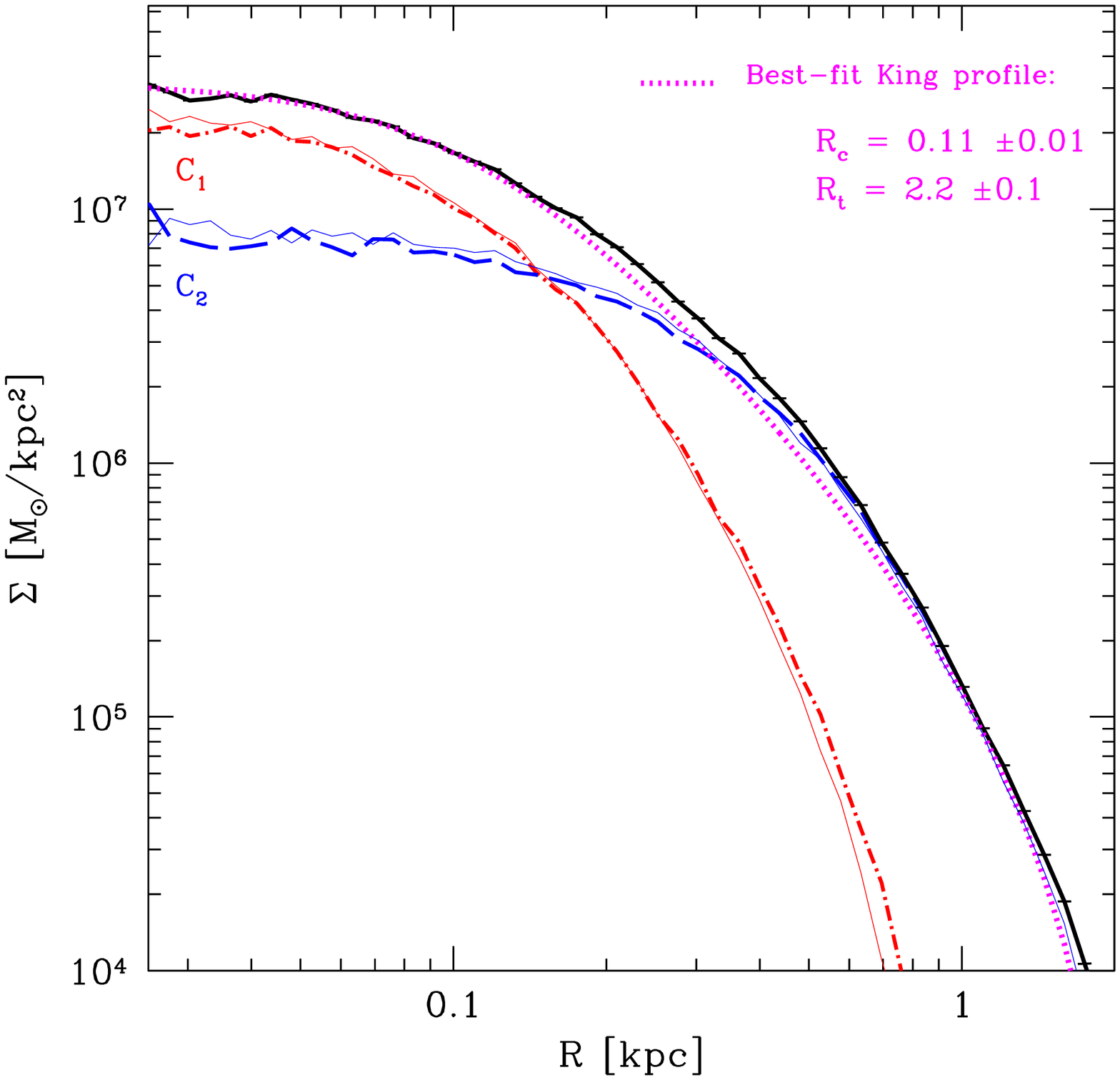}
\includegraphics[width=0.485\linewidth]{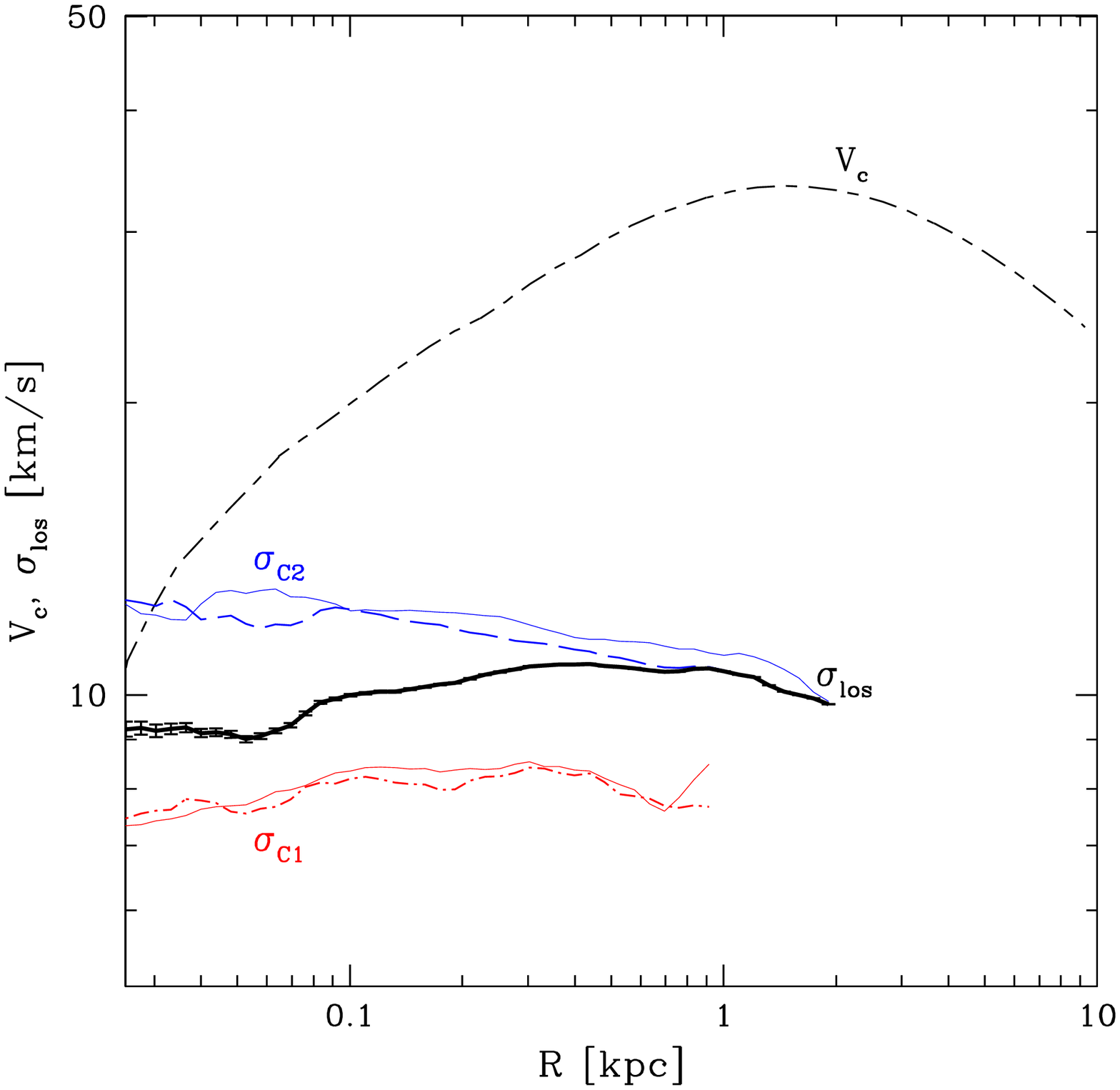}
\caption{Surface mass density (left) and velocity (right) profiles for
  the simulated satellite as a function of the (projected) distance
  from the centre, $R$. In both panels, red dot-dashed and blue dashed
  curves distinguish the contributions from the more concentrated
  ($C_1$) and more extended ($C_2$) components respectively, while the
  thick solid black shows the total values obtained for all the stars
  (i.e. no distinction between populations).  In the left panel, the
  magenta dotted line shows the best-fit King profile obtained for the
  total distribution of stellar mass, and the numbers quoted
  correspond to the derived core and tidal radii, respectively. The
  right panel shows the circular velocity (black dashed) plus the
  line-of-sight velocity dispersion profiles associated to each
  luminous component. Poisson uncertainties are shown for the total
  density and velocity dispersion profiles in our model, although they
  are generally smaller than the thickness of the curves. Thin lines
  in both panels correspond to a control model, which is evolved from
  this initial configuration in isolation during $t \sim 6$ Gyr.}
\label{fig:rho}
\end{center}
\end{figure*}

The softening lengths in multiple component systems must be chosen
carefully. We follow the prescriptions described in
\citep{athanassoula00}, which give in each case are: $\epsilon_{C1}
\sim 7 \rm pc$, $\epsilon_{C2} \sim 21 \rm pc$ and $\epsilon_{DM} \sim
60 \rm pc$. Figure \ref{fig:rho} shows the resulting (projected)
density and velocity profiles for this model (hereafter, $R$ and $r$
refer to projected and three-dimensional distances, respectively).  As
a consequence of this set-up, the initial configuration of the
simulated dwarf has a slightly rising $\sigma_{\rm los}$ profile in
the inner regions, that can be explained as the transition from the
more concentrated to the more extended stellar components \citep[see
][]{mc07}.

The baryonic components of the simulated dwarf are deeply embedded
within the potential well of the dark matter halo: the core radius is
$\sim 7.3\%r_{\rm max}$, where $r_{\rm max}$ is the radius at which
the circular velocity of the dark matter halo peaks; in good agreement
with suggestions from numerous theoretical models
\citep{lokas01,hayashi03,penarrubia08a}.
Such segregation with respect to the dark matter halo turns the
stellar components of the satellite galaxies more resilient to tidal
stripping, increasing the probability of survival for many orbital
times without appreciable changes in their observable properties.
Notice that stars account only for less than 0.25\% of the total mass,
the bulk of the dwarf mass is largely dominated by the dark matter
component, with a $\rm M/\rm L \sim 35$ within the tidal radius (we
have assumed here a conversion factor $\gamma=2.9$ to compute the
luminosity out of the mass in the baryonic components, this is roughly
consistent with a $\sim 10$ Gyr old stellar population of mean
metallicity [Fe/H]$=-2$ (as derived from the BASTI model
isochrones)\footnote{www.te.astro.it/BASTI/index.php}.

The stellar and dark matter masses have been chosen to approximately
match the typical line of sight velocities dispersions ($\sigma_{\rm
los} \sim 10$ km/s), luminosities and structural parameters (such as
core $R_c$ and tidal $R_t$ radius) that are observed in Local Group
satellites. Throughout this paper we define the characteristic radii
$R_c$ and $R_t$ by finding the best-fitting King profiles to the projected
density of particles in the luminous components. Notice that with
this definition, the tidal radius is not necessarily related to the
physical radius that divides the bound from the unbound population of
stars in cases where tidal stripping takes place.

The simulated satellite constructed in this way is first relaxed
(evolved in isolation) during 1 Gyr before being placed on orbit around the
host potential for a longer timescale of  $t \sim 6$ Gyr.

  The stability of our initial conditions against numerical and
  relaxation effects can be appreciated in Figure
  \ref{fig:rho}. Overplayed using thin lines we show the final
  configuration (projected density and velocity dispersion profiles)
  for the same satellite model but evolved in {\it isolation} during
  $t \sim 6$ Gyr.  The close match between the initial conditions
  (thick lines) and this control model (thin lines) indicates that any
  departure from the initial set up will be driven by the evolution of
  the satellite into the host potential and is not a consequence of an
  out-of-equilibrium initial configuration.

\subsection{Satellite Orbits}

\begin{figure}
\begin{center}
\includegraphics[width=0.5\linewidth]{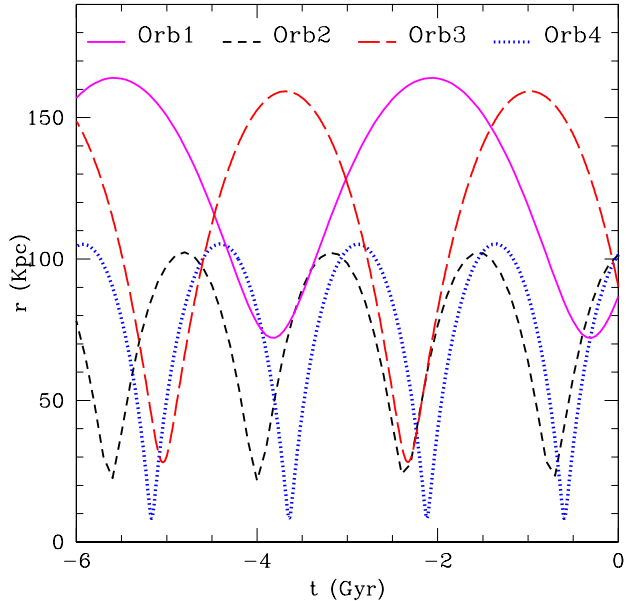}
\caption{ Galactocentric distance $r$ as function of time for the four
different orbits explored in this article.  Orbit 1 and 2 have been generated 
from the present$-$day positions
and proper motions of the Sculptor and Carina dwarfs, as reported by
Piatek et al. (2003,2006) respectively.}
\label{fig:orbits}
\end{center}
\end{figure}

The gravitational forces that a satellite galaxy experiences are
strongly dependent on its orbit, in particular on the pericenter
distances and the orbital radial period (that determine the number of
pericenter passages that occur in a given time span), both of which are
related to the eccentricity of the orbit.  While the dark matter halo
associated to a satellite provides some sort of ``shielding'' for the
baryons, the removal of the luminous component may start after the
dark mass has been reduced to less than 10\% its original value.
However, the changes in the equilibrium state produced by a
significant depletion of the dark matter halo could in principle
affect the properties of the baryonic component even before begin
to be stripped.

We explore this by placing our fiducial satellite in four different
orbits within the host potential (see Figure \ref{fig:orbits}), and
integrating its evolution in time for $t \sim 6$ Gyr (this may vary
$\pm 0.3$ Gyr depending on the timing of their orbit apocenter
passage). This integration timescale is chosen to be consistent with
the average time of accretion of the surviving population of
satellites found in semi-analytical models and numerical simulations
of galaxy formation
\citep{bullockandjohnston05,font06a,font06b,sales07a}. In order to
facilitate the interpretation of the results in terms of Local Group
galaxies, we have generated two orbits from the present day positions,
radial velocities and proper motions of Sculptor and Carina, as
reported by Piatek et al. (2003,2006) \citep{piatek03,piatek06}. These
correspond to Orb1 and Orb2 in Figure \ref{fig:orbits}, for Sculptor
and Carina respectively.  Even though the errors in the velocity
determinations are appreciable ($\sim 50\%$), which leads to a wide
range of possible orbits, the more likely values give rise to
interesting differences in their orbits (Orb1 and Orb2). Orbit 3 and
4 correspond to two random orbits generated under the condition
of comparable present-day galactocentric distances ($r \sim 90-100 \rm
kpc$) to orbits 1 and 2.

Figure \ref{fig:orbits} shows that despite their similar present-day
distance from the host, an orbit such as that suggested by Sculptor's
proper motions (Orb1) is significantly more external than Carina's
(Orb2), with apocenters exceeding 150 kpc and pericenter distances of
$\sim$75 kpc. Orb2 is much more constrained to the inner parts of the
Milky Way potential, with pericenters smaller than 30 kpc and orbital
periods $\sim 1.6$ Gyr, almost 2.4 times shorter than estimated for
Scl. Orb3 has comparable pericenters to Orb2, but its apocenters are
50\% larger, which translates into a longer orbital radial
period. Finally Orb4 is similar to Orb2, and is also consistent with
Car proper motions, but has much smaller pericenter values
($r_{per}\sim 9$ kpc).

A given satellite placed on these four orbits will thus be subject to
different tidal field strengths, and consequently, will present
structural and dynamical properties that may differ substantially
after $t \sim 6$ Gyr of evolution in the host potential. We focus on
these aspects in the next Section, with special attention to the
effects imprinted in the internal kinematics of the stellar
components. At this point in time, the satellite is located near
  the apocenter for Orb2 and Orb4, where it will spend most of its
  time. The system will then re-configure to its new equilibrium state
  and the configuration during apocenter passages is the most
  representative of their final state. On the other hand, at the final
  snapshot, satellites in Orb1 and Orb3 are at arbitrary points on
  their orbits. However, given their more external orbital
  configuration, the final properties for these systems do not
  strongly depend on the location along their orbits.

\begin{figure}
\includegraphics[width=0.95\linewidth]{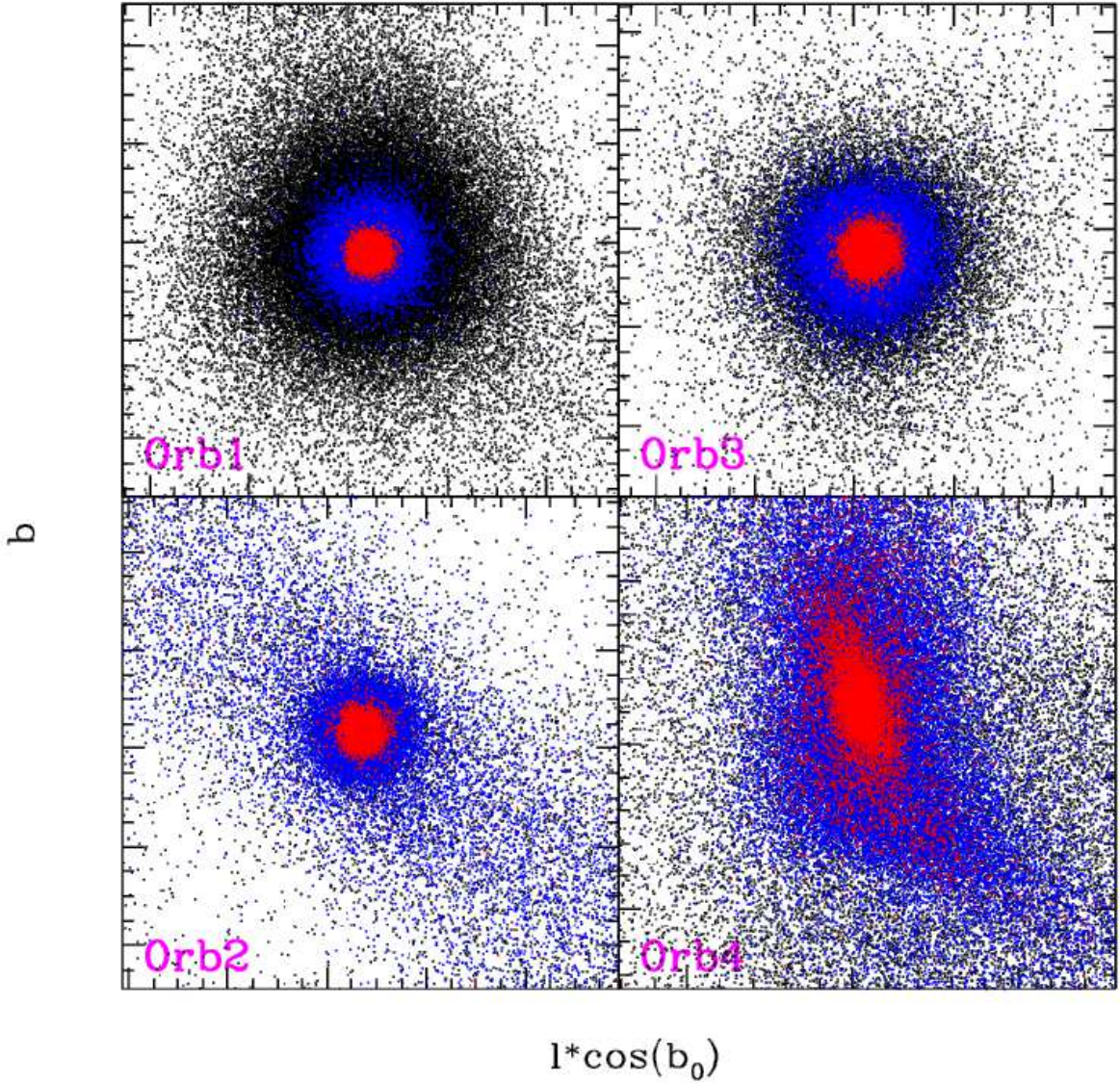}
\caption{Final projected view (arbitrary units) of the dark matter (black), metal poor (blue) and
metal rich (red) stellar components for the models evolved in orbits 1-4 during 
$t \sim 6$ Gyr. 
Due to the different degree of tidal disruption experienced in each case, 
some satellites have turned more elliptical in shape in either one (Orb 2) or 
both (Orb4) their stellar components. Satellites in orbits 1 and 4 remain still
quite spherical as the progenitor model.
}
\label{fig:shapes}
\end{figure}

\section{Results}
\label{sec:results}

After $\sim 6$ Gyr of evolution of the model satellite in the host
potential, we are able to identify a self-bound satellite remnant for
all four explored orbits. However, in spite of the originally
identical settings, the structural properties of the surviving dwarf
vary appreciably depending on the different orbits. A brief summary of
the retained mass fractions and the structural parameter for each model 
can be found in  Table \ref{tab:table1}.

\subsection{Final distribution of satellite's stars}

Figure \ref{fig:shapes} shows a snapshot view of the final system
  evolved in each of the orbits. Black, blue and red correspond to the
  dark matter, metal poor and metal rich stellar components,
  respectively. Visual inspection of this Figure suggests that the
  shapes of the remnants at the final time keep some imprints about
  the history of each dwarf. For example, for external orbits such as
  Orb1 and Orb3, the stars in both components are distributed in a
  spherical fashion, reminiscent of (and determined by) the initial
  setting of the progenitor.  On the other hand, satellites on Orb2
  and Orb4 have developed tails of unbound material as a result of a
  stronger tides suffered during their small pericenter passages. It
  is interesting to note that since tidal disruption affects earlier
  and more appreciably the extended $C_2$ component than $C_1$, the
  shape of the metal poor stellar distribution may be more elongated
  than the metal rich when mass depletion has not yet
  proceeded into the inner regions of the dwarf. Such is the case of
  the model in Orb2, which has retained 100\% and 88\% of the initial
  mass in components $C_1$ and $C_2$, respectively (see Table
  \ref{tab:table1}). However, once tidal stripping affects also the
  more concentrated population, the distribution of metal rich and
  metal poor stars will both turn elliptical in shape as seen for Orb4.

\begin{figure}
\includegraphics[width=0.95\linewidth]{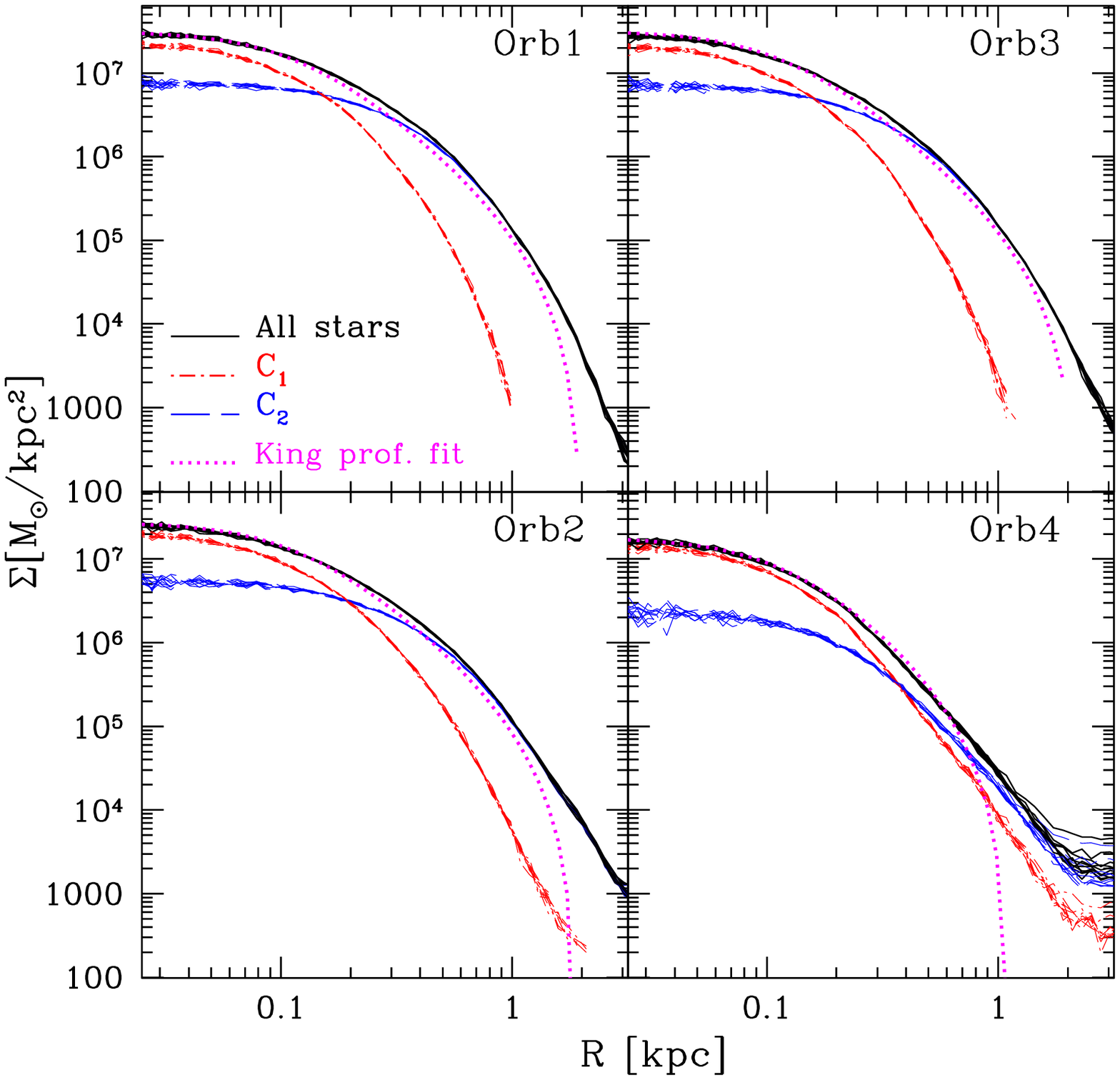}
\caption{Surface mass density profiles of all our models after $\sim 6$
 Gyr of integration in the host potential. The color codding is the same
 as used in Figure \ref{fig:rho}}
\label{fig:rho_all}
\end{figure}

The final projected mass density profiles for each simulation are
shown in Figure \ref{fig:rho_all}, together with their best-fitting
King profiles.  The corresponding core and tidal radii ($R_c$ and
$R_t$ respectively) are listed in Table \ref{tab:table1}, together
with the fractions of mass in each component that has remained bound.
We have tested the effect of the particular alignment of the
  tidal tails with the line of sight to the observer by computing the
  density profiles obtained from 10 different random projections of
  each satellite. These are depicted by thin lines in Figure
  \ref{fig:rho_all}, which show that orientation is not particularly
  important for the fairly undisturbed systems, and its impact is
  minor in the case of Orb4.

\begin{table}
  \begin{center}
    \begin{tabular}{|c|c|c|c|c|c|}
      \hline \hline
      Orbit & $R_c$ (kpc) & $R_t$ (kpc) & $f_{C1}$ (\%) & $f_{C2}$ (\%) & $f_{DM}$ (\%)\\
      \hline
      Initial  &  $0.11\pm0.01$ & $2.2\pm0.1$ & 100 & 100 & 100 \\
      Orb1 &  $0.11\pm0.01$ & $2.0\pm0.1$ & 100 &  99 & 44 \\
      Orb2 &  $0.11\pm0.01$ & $1.9\pm0.1$ & 100 & 88 & 5 \\
      Orb3 &  $0.11\pm0.01$ & $2.2\pm0.1$ & 100 & 98 & 20 \\
      Orb4 &  $0.11\pm0.01$ & $1.0\pm0.1$ & 87 & 20 & 3 \\
      \hline
    \end{tabular}
    \end{center}
    \caption{Structural properties of the satellite models after its evolution for
      $t =6 (\pm0.3)$ Gyr within the fixed Galactic potential.  The first
      row in the Table shows the initial values and subsequent
      rows show the present-day values for the experiments on Orb1-Orb4. $R_c$ and
      $R_t$ refer to the best-fitting King profile values to the
      surface density profile of the luminous components.  The
      last 3 columns indicate the fraction of mass in each
      component that remains bound to the dwarf in the final
      configuration.}
  \label{tab:table1}
\end{table}

As quoted in Table \ref{tab:table1}, the tidal forces have affected
significantly less the stellar components than the associated dark
matter halo of the dwarf. For example, the satellite on orbits 2 and 3
has retained only 5\% and 20\% of its dark matter mass respectively,
but more than 90\% of the initial stellar mass. Notice that the
structural parameters of the baryons (quantified by the projected King
profile best fit parameters) are robust to the ongoing pruning of the
satellite's outskirts, in particular, the core radii show no
variations in any of the experiments \citep[see discussion in
ref.][]{penarrubia08b}. On the other hand, the external parts, and
consequently the tidal radius do show more significant changes. In the
case of the orbit with the smallest pericenters (Orb4) the tidal
radius has shrunk by a factor $\sim 2$ from its original value. 
Also projection effects are in this case more important, due to
different contributions of unbound stars according to the alignment of
the tails with the line of sight \citep{klimentowski07}. For the
other three orbits, the tidal radius and density profiles have been
only barely affected by the stripping.

We observe that the segregation in the spatial
distribution between components $C_1$ and $C_2$ is also modified,
although remains present in all of our experiments (however,
this clearly depends on the length of the integration, 
see Section \ref{sec:longer_time}). For example, the
mass ratio between $C_1$ and $C_2$ within the core radius tends to
increase in cases where tidal stripping is significant. This is
naturally expected as tides will lead to a larger (and earlier)
stripping of the most extended ($C_2$) component compared to
$C_1$. For orbits 2 and 4, $m_{C_1}/m_{C_2}$ increases from $\sim 2$
in the initial model, to 2.6 and 4.9, respectively (orbits 1 and 3
show no variation). For the same orbits (2 and 4) this ratio for the
more external region $0.5 < R <1 \rm$ kpc changes from an initial
value $m_{C_1}/m_{C_2}\sim 0.02$, to $\sim 0.12$ and $\sim 0.62$,
respectively.

This gradient of $m_{C_1}/m_{C_2}$ with radius implies that the
spatial distinction between both luminous components is still present
in the dwarf galaxy. Notice however that some redistribution of the
stars in the components has taken place.  For example, in the external
regions ($0.5< R <1$ kpc) of the dwarfs in orbits 2 and 4, the
contribution of particles from component $C_1$ changes from negligible
(less than $\sim 2\%$) to significant ($\sim 10-60\%$) as a result of
the ongoing tidal perturbations that require the dwarf to find new
equilibrium configurations (see also Figure \ref{fig:shapes}).

\begin{figure}
\includegraphics[width=0.98\linewidth]{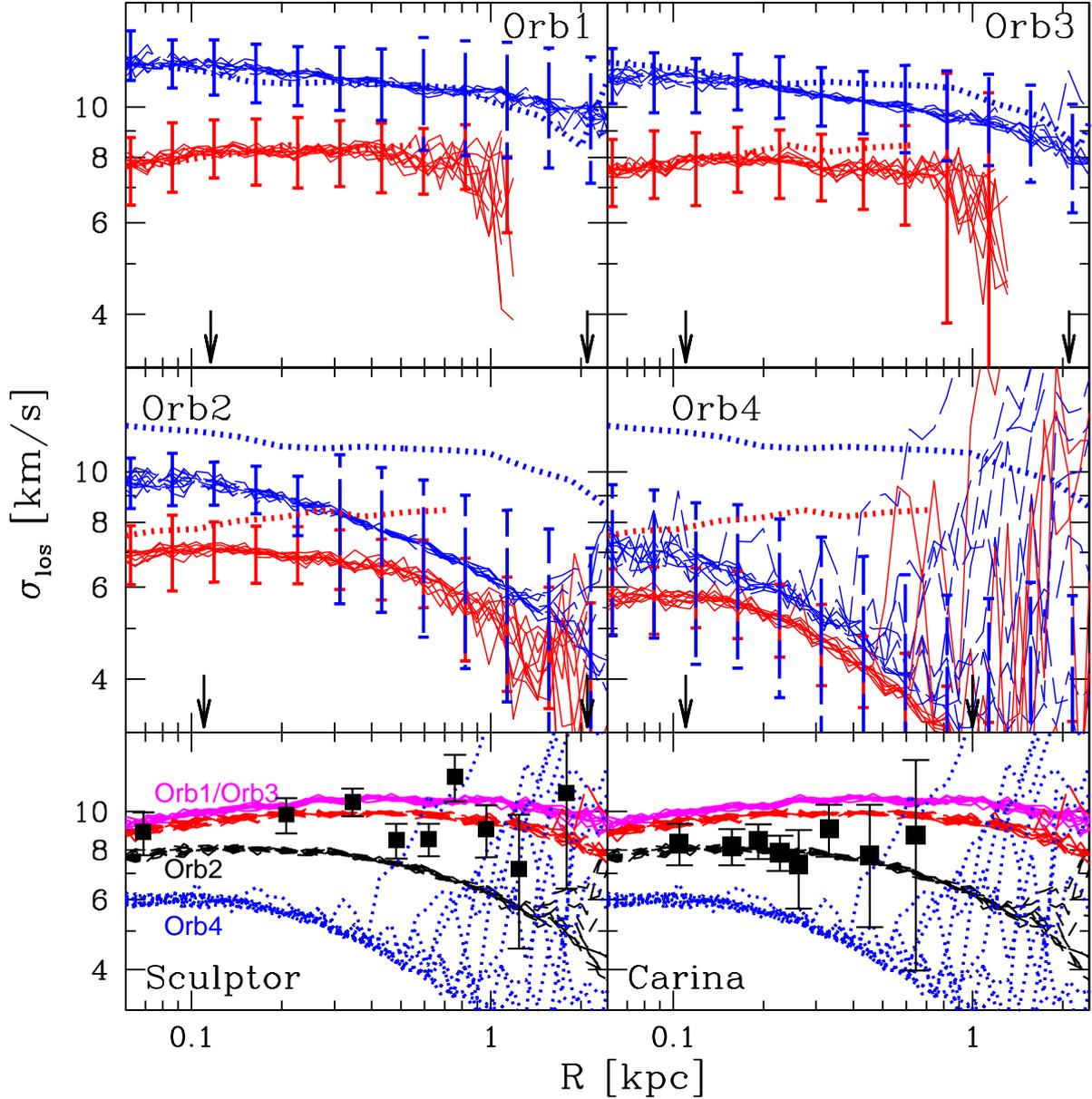}
\caption{{\it Upper four panels:} Line of sight velocity dispersions
  for both luminous components as a function of the (projected)
  distance from the center of the satellite. Solid red and dashed blue
  curves correspond to the concentrated ($C_1$) and the more extended
  ($C_2$) components, respectively. The error bars indicate the rms
  dispersion on the $\sigma_{\rm los}$ profiles of 100 random samples
  containing 500 particles each (see text for details). For reference,
  the dotted lines show the initial configuration of the model
  satellite. Black arrows indicate the positions of the core and tidal
  radii in each experiment.  {\it Bottom two panels:} Velocity
  dispersion profiles for Sculptor and Carina dwarfs (squares with
  error bars).  Data was taken from Battaglia et al. (2008) and Koch
  et al. (2006), Helmi et al. (2006)
  \citep{battaglia08,koch06,helmi06}.  The {\it total} (no distinction
  between components) $\sigma_{\rm los}$ profiles for our models in
  the 10 different projections are also shown in magenta, black, red
  and blue lines for Orb1-Orb4, respectively.}
\label{fig:sigma_rp}
\end{figure}

\subsection{Dynamics of satellite's stars}

Also worth highlighting is the dynamical response of the system to
such structural changes.  The velocity dispersion in spherically
symmetric systems is determined by the amount of mass contained within
a given radius (i.e. the density profile) and by the velocity
anisotropy associated to the particles orbits
\citep{binneyandtremaine08}. We naively expect from the results
discussed above that not only the structural parameters but also the
kinematical properties (such as the line-of-sight velocity dispersion,
$\sigma_{\rm los}$) will be affected by the mass removal experienced
by the satellite.  This is shown in Figure \ref{fig:sigma_rp}, where
we plot the (1D, line-of-sight) velocity dispersion profiles as a
function of the projected radius from the center of the satellite. To
avoid contamination by unbound particles, and following the method
applied to observations \citep{battaglia08}, we have removed all
particles whose line-of-sight velocity exceeds by $\pm 3 \sigma_{\rm
  los}$ the mean velocity of the system ($|\bar{V}(R)-V_{\rm
  los}|>3\sigma_{\rm los}$).

The panels in Figure \ref{fig:sigma_rp}
correspond to each one of the orbits explored, with thin lines
showing the velocity dispersion in each radial bin for each
population. The long dashed blue and the solid red curves refer to the
more extended ($C_2$) and concentrated ($C_1$) luminous components,
respectively. In order to illustrate the effects introduced by the
geometry of the problem and the different alignments between possible
streams and the line of sight direction we show curves corresponding to 10 
different random projections for each satellite. 
The error bars in this Figure are estimated as
follows. We randomly extract a (sub)set of 500 particles from the
luminous components of the simulated satellite, maintaining the mass
ratio between population $C_1$ and $C_2$ and their spatial
distribution. The size of each subsample (500 points) is comparable to
the size of current observational samples of line-of-sight velocities
in dSphs \citep{battaglia08,walker07}.  For each subsample we derive
$\sigma_{\rm los}$-$R$ curves, similar to those shown in Figure
\ref{fig:sigma_rp}, albeit with larger noise due to the poorer number
statistics. The average scatter between these $\sigma_{\rm los}$-$R$
relations obtained from the 100 realizations is then shown as error
bars in Figure \ref{fig:sigma_rp}.  In order to guide the eye, we have
included as well in this Figure the initial conditions for the
satellite (dotted lines).

Figure \ref{fig:sigma_rp} shows that when a strong segregation exists
between the luminous and dark components, the velocity dispersions of
the stars are only weakly affected by the removal of the extended dark
halo. For example, experiments Orb1 and Orb3 show that the dark matter
halo has been reduced to only $\sim 45\%$ and $20\%$ respectively of the
initial mass, yet the kinematics of the luminous components remained
essentially intact.  {\it Tidal effects proceed outside-in, preserving
the central regions of the satellites (where baryons are located)
roughly unchanged}.

However, the situation is different when trimming sets in on the
stellar components, as is the case for the satellite on orbits 2 and 4
(bottom panels in Figure \ref{fig:sigma_rp}). The tidal stripping tends
to shrink the initial gap between the velocity dispersion of the
luminous components. Furthermore, the limited sizes of available
observational samples ($\sim 500$ member stars) will lead to
additional uncertainties preventing a clear distinction between the
kinematics of the two populations \citep[see ][]{ural08}.  Even though
components $C_1$ and $C_2$ experience a decrease in their $\sigma_{\rm
los}$, they are not affected in the same degree, mainly due to their
relative spatial segregation that determines an earlier and stronger
removal of stars from component $C_2$ compared to $C_1$. In such
cases, the preferential stripping of the more extended component
(mimicking the metal-poor component of satellites) should lead to the
presence of metallicity gradients along the stellar streams of the
remnant. Evidence for such trends have recently been found in the
streams of the Sagittarius dwarf, where BHB are significantly (at
$\sim 4.8\sigma$ level) more abundant than red clump stars in
comparison to the core of this galaxy \citep{bellazzini06}.  This is
interpreted as the preferential removal of the old metal poor (BHB)
stars from the outskirts of Sagittarius, compared to the
intermediate$-$age red clump population that has remained in the core
of the dwarf less affected by external tides, in strong analogy with
our simulations.

The velocity dispersion profiles for each of the luminous components
decline with radius (instead of being roughly flat as in the initial
model) as stripping proceeds on the simulated galaxies. Beyond the
tidal radius (recall this is defined purely by the best-fitting King
profile to the projected mass distribution of the stars), the velocity
dispersion increases reaching $\sigma_{\rm} \geq 15$ km/s, as a result
of the larger contribution from gravitationally unbound stars (that
the $\pm 3\sigma$ cut described above is unable to remove). This
  effect is particularly important for Orb2 and Orb4, where the tidal
  stripping of stars proceeded. Note that this upturn in
  $\sigma_{los}$ disappears once the analysis is restricted to
  particles that remain gravitationally bound to the satellite.  In
  agreement with Klimentowski et al. (2007) \citep{klimentowski07}, we
  find that depending on the orientation of such tails with respect to
  the line of sight, different $\sigma_{\rm los}$ profiles are
  obtained. When the radial vector to the satellite is well aligned
  with the direction of these tails, the contamination by unbound
  material is maximized, producing a raising velocity dispersion
  profile towards the dwarf's outskirts.

Taken at face value, these velocity dispersion profiles may seem at
odds with the typical flatness of the $\sigma_{\rm los}$ profiles
observed for Local Group dwarfs. As an example, observed data
  corresponding to the line of sight velocity dispersion inferred for
  Sculptor and Carina are shown in the two bottom panels of Figure
  \ref{fig:sigma_rp}. Data for these two dwarfs was taken from
  \citep{battaglia08} and \citep{koch06,helmi06}, respectively.
However, unless the stripping is considerably large, our simulations do
still reproduce reasonably flat (or even rising) {\it total} velocity
dispersion profiles, when no distinction between stars from population
$C_1$ and $C_2$ is made \citep[see also ][]{mc07}.  This is shown
  by the magenta, black, red and blue lines in bottom panels of Figure
  \ref{fig:sigma_rp}, corresponding to the total velocity dispersion
  profiles predicted for Orb1-Orb4 (notice color-codding follows
  Figure \ref{fig:orbits}).  Our models in Orb1 and Orb2 reasonably
  fit the observed velocity dispersion data for Sculptor and Carina,
  respectively.

We find that while projections have profound effects on $\sigma_{\rm los}$, 
it only causes mild variations on the projected
density profiles around the formal tidal radius (see Figure
\ref{fig:rho}). A closer look at this indicates that while only $\sim
5-10\%$ contamination from unbound stars is enough to increase the
velocity dispersion profiles by few km/s, at least twice this
fraction is needed to drive significant changes on the surface density profiles. 
For example, the model in orbit 4 shows an increase in
$\sigma_{\rm los}$ at $R \sim 600$ pc, where the fraction of unbound
stars accounts for $\sim 8-10\%$ of the mass at that radius. However,
bottom right panel of Figure \ref{fig:rho} shows that an excess of
stars appears in the surface density only for $R > 1500$ pc, when
the unbound contribution is $\geq 20\%$.

We finally point out that the shape of the line of sight velocity profiles 
may depend on the level of rotational support of the initial progenitor, 
which we have neglected in this study. A comparison between properties of the
remnant of a disky versus a spheroidal progenitor is undoubtedly interesting
and deferred to future work.

\begin{figure*}
\includegraphics[width=0.471\linewidth]{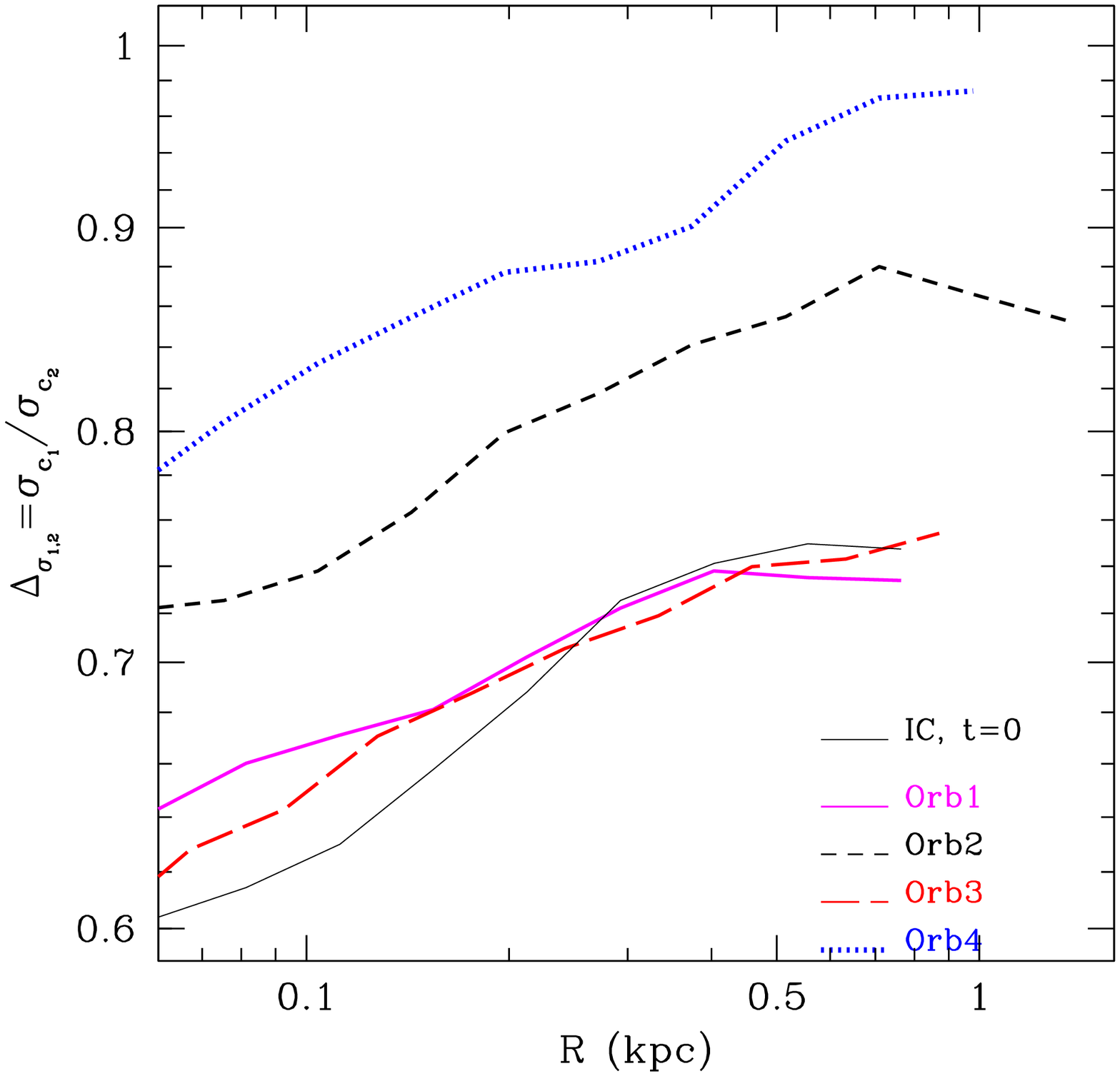}
\includegraphics[width=0.471\linewidth]{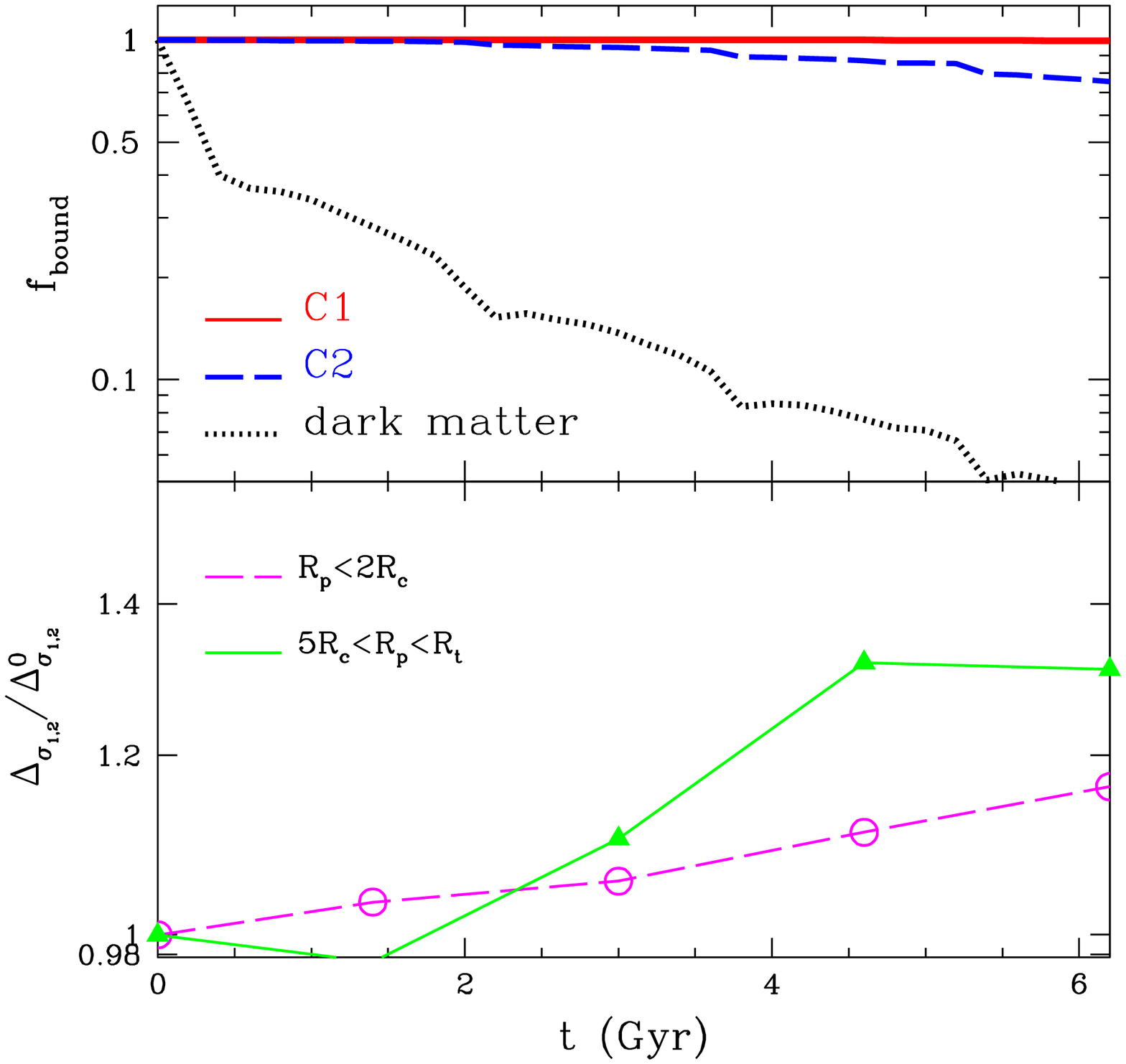}
\caption{ {\it Left:} Ratio between the line of sight velocity
  dispersions for component $C_1$ and $C_2$
  (${\Delta_{\sigma_{1,2}}}=\sigma_{c1}/\sigma_{c2}$) as a function of
  the projected distance from the center of the dwarf. The mos distant
  bin considered has at least 20 particles from each component (and more for
  for bins closer in). {\it Right:} The upper
  panel shows the fraction of mass (compared to the initial value)
  that remains bound to the satellite as a function of time for orbit
  2. Dotted, red solid and blue dashed curves are used to
  indicate the dark matter, and stellar components $C_1$ and $C_2$
  respectively. In the bottom the evolution with time of the ratio
  between the velocity dispersions of $C_1$ and $C_2$ are shown
  for two regions within the dwarf: stars within two core radii
 (magenta dashed) and stars approximately lying closer to the
  tidal radius (green solid). These ratios have been normalized to the
  original values (${\Delta_{\sigma_{1,2}}}^0$, thin black line in the
  right panel).}
\label{fig:deltasig}
\end{figure*}

The preferential pruning by tides of the more extended component is
the main responsible for the attenuation of the kinematical
distinction between both stellar populations. This can be seen on the
left panel of Figure \ref{fig:deltasig}, which shows the final
velocity gap between components $C_1$ and $C_2$ as a function of
distance from the center of the satellite. The vertical axis
corresponds to the ratio between the velocity dispersion in these
components, $\Delta_{\sigma_{1,2}} = \sigma_{C_1}/\sigma_{C_2}$, for
each of the orbits we have explored. For an easier comparison, the
black dotted curve shows the initial configuration.  When the model
satellite is evolved on orbits such as 1 or 3 (solid magenta,
long-dashed red) the distinction set by construction between both
luminous populations remains essentially unchanged at any radius
within the dwarf. In spite of the pruning of 60-80\% of the dark
matter content on these satellites, these two experiments showed
negligible stripping of their stellar components, signaling that a
certain degree of stellar mass removal must take place before the
relative kinematics of the two populations is noticeably altered.

On the other hand, the simulated satellites on orbits 2 and 4
experienced significant stellar stripping and show kinematical gaps
between the luminous components that have been reduced by $\sim 20-30\%$.
Such ``mixing'' proceeds typically outside-in,
starting soon after the tides begin to strip stars from the dwarf. The
right panel of Figure \ref{fig:deltasig} shows the evolution of the
bound mass with time (upper panel) together with the kinematical gap
(bottom panel) for the simulated satellite on orbit 2. Notice that
mass removal is not a continuous process but rather presents a
stepwise behaviour closely related to the orbital pericenter crossings. After
the first pericenter passage, the dark matter content of the satellite
has dropped by almost a factor of 2. However, the stellar mass is more
resilient and shows signs of ongoing stripping only after the third
pericentric passage, when the dark matter mass has shrunk to only $\sim 10\%$
of the original value.  It is thus when the stripping of the stellar
mass begins that the relative kinematical differences between the two
luminous components show a significant departure from the original
configuration (see bottom right panel of Figure \ref{fig:deltasig}).  This
figure also shows that only the more extended ($C_2$) stellar
component is affected, with $C_1$ conserving 100\% of its initial mass
during the whole time integration. This helps to explain why the
kinematical gap changes more appreciably in the outskirts than in the
core of the satellite galaxy, which will tend to remain less affected
by tidal forces.

\subsection{Effect of a different time integration}
\label{sec:longer_time}

 Models that attempt to reproduce the present-day properties of
  dwarf galaxies are usually degenerate with respect to the unknown
  initial structure of the objects (density profile, mass,
  characteristic scales) and their orbital paths.  Even though current
  data allow to reasonably characterize the present-day structural
  properties of the nearby dwarf galaxies, these are a combined result 
  of: $i)$ their initial configuration,  $ii)$ their orbital motion within the
  Local Group and $iii)$ evolution of the host gravitational potential field; 
  all three factors with hardly any strong observational
  constraints.  Proper motions help us to partially reconstruct the
  orbits, but unfortunately, errors are still prohibitive, with
  uncertainties of the order $\sim 20-100\%$
  \citep{piatek03,piatek06,walker08}.

\begin{figure*}
\includegraphics[width=0.471\linewidth]{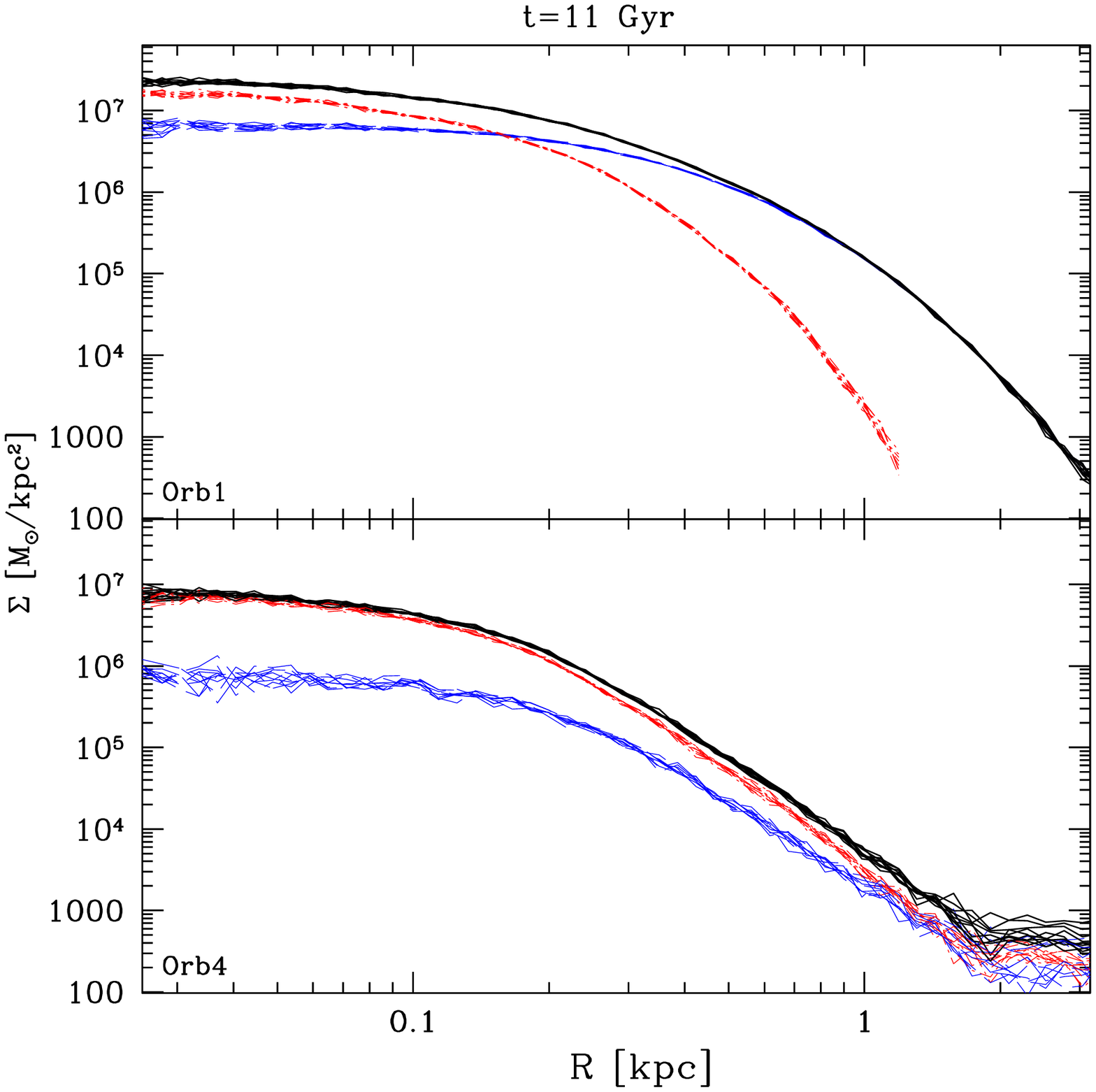}
\includegraphics[width=0.471\linewidth]{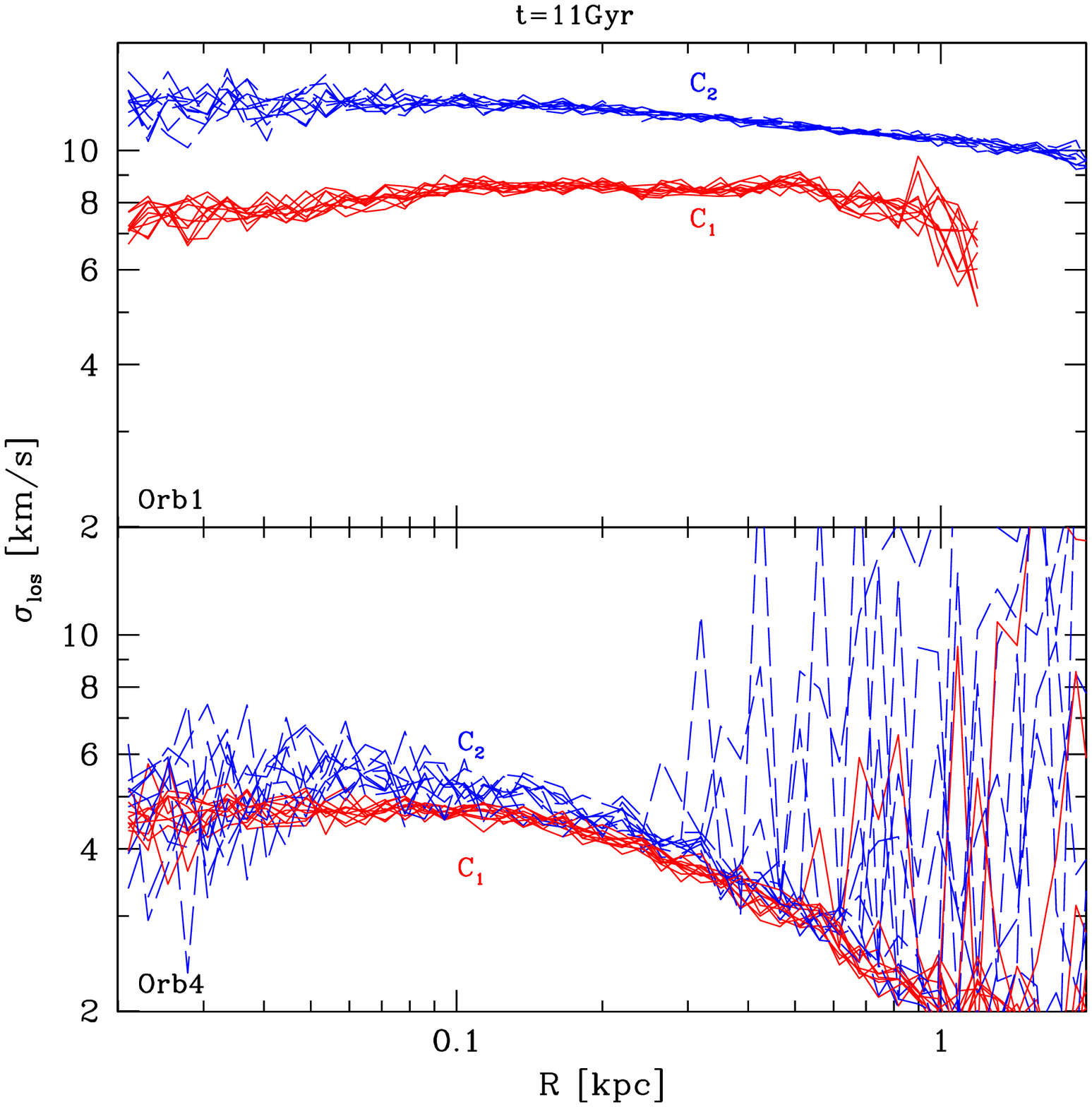}
\caption{Projected density (left) and velocity dispersion (right) profiles
for satellites evolved in Orb1 and Orb4 for a longer time interval: $t \sim 11$ Gyr,
instead of the fiducial $t \sim 6$ Gyr. Notice for Orb4 that, as mass depletion proceeds
further, the spatial and kinematical segregation between $C_1$ and $C_2$ are
increasingly erased.
}
\label{fig:long_t}
\end{figure*}

  As a rule-of-thumb, for a fixed final configuration, more massive,
  concentrated objects are needed if we desire to increase the time
  integration in the host potential keeping the orbit of the object
  fixed \citep[e.g. see Section 3.3 in Ref. ][]{sales08}.  For the analysis
  presented above, we have chosen a fiducial value for the time
  integration of $t \sim 6$ Gyr; which is, at some extent, arbitrary. We
  explore in Figure \ref{fig:long_t} the effects induced by a
  longer time integration ($t \sim 11$ Gyr) on the density profile
  (left) and velocity dispersion (right) for two of our models (Orb1
  and Orb4).

  For external orbits such as Orb1, with long orbital period and large
  pericenter distances, the effect of (almost) doubling the time
  integration is negligible. Objects on such orbits can preserve most
  of their luminous component bound, showing very little evolution in
  their spatial and kinematical properties with respect to the initial
  configuration.

  On the other hand, if tides are as strong as on Orb4, the increase
  in the integration translates into lower densities in the core of the
  dwarf, and to a further mixing of the spatial and kinematical
  profiles of both stellar populations. For example, for the satellite
  evolved in Orb4 during $t \sim 11$ Gyr, the bound mass fraction of
  components $C_1$ and $C_2$ drop to about 38\% and 4\% of their
  initial configuration, respectively. This seems to be enough to completely
  erase the kinematical gap between both components (see bottom right
  panel of Figure \ref{fig:long_t}) as well as their initial spatial
  segregation (bottom left panel).

  We note that this object shows, at the end of the experiment, an
  overall velocity dispersion $\sigma \sim 5$ km/s, that is about half
  the initial value; quite low compared to the typical value of the
  classical satellites in the Local Group. However, given the
  degeneracies in this type of modelling mentioned above, a larger
  velocity dispersion $\sigma_{los} \sim 8-9$ km/s may be reconciled
  with a time integration as long as $t \sim 11$ Gyr provided we
  increase accordingly the initial $\sigma_{\rm los}$ of the
  progenitor system \citep{sales08}.

  The behaviour shown in Figure \ref{fig:long_t} confirms the
  conclusions from Section \ref{sec:results}; namely that tides
  affecting the luminous components are needed in order to effectively
  erase the structural and kinematical biases between two initially
  segregated stellar populations.

\section{Application to the Local Group: Sculptor and Carina dwarfs}

The results reported in Section \ref{sec:results} can be extrapolated to
particular examples within the Local Group. Among the dwarf
spheroidals orbiting around our Galaxy, Carina and Sculptor are
perhaps the most instructive cases to interpret in light of our
results given their proximity and good quality data currently
available.  Sculptor shows clear spatial variations in the properties
of its stellar populations, such as spatial variations in the relative
distribution of RHB and BHB stars, a clear metallicity variation with
radius as derived from spectroscopic data and also kinematical
differences associated to each metallicity population.  Carina instead
shows no gradient in the distribution of horizontal branch stars
themselves, but a distinction is possible between HB stars and red
clump stars. Signs of a weak metallicity gradient was found by Koch
et al. (2006) \citep{koch06} using $\sim$500 giant branch
stars. However this gradient is only mild and, unlike Sculptor, no
kinematical distinction has been identified so far.

The Sculptor and Carina dSphs are found today at comparable distances
from the Milky Way ($\sim 100$ kpc), although numerical integration of
their proper motions suggests that their orbital paths could have been
quite different in the past (see Orb1 and Orb2 in Figure
\ref{fig:orbits}). The most likely orbit of Carina seems to be
confined much more to the inner regions of the Milky Way halo, with
smaller pericenter passages and shorter periods than that of
Sculptor. Our numerical experiments suggest that a satellite galaxy
orbiting in a Carina-like orbit (Orb2) is likely to be exposed to
tidal forces that drive significant mass removal. This in turn
triggers a mixing and weakening of the kinematical gap that may have
existed between its luminous components, which may be challenging
to detect with the current spectroscopic samples.

On the other hand, the same model evolved in a more external orbit
like Sculptor's maintains the kinematical and spatial properties of
the luminous components almost unchanged after $\sim 6$ Gyr of
evolution in the host potential.

A simple extrapolation of these results suggests that the lack of a
stronger and clear metallicity gradient in Carina, or equivalently,
the non-detection of distinct kinematics associated to different
stellar populations might be partly due to the effect of strong tidal
forces inherent to Carina's orbit. It is possible that in the past,
Carina could have had a double component configuration, where the
metallicity gradient and kinematical gaps were larger than today's
measurements suggest.  Notice that according to our simulations, this
can only happen if tidal stripping has already taken place on the
luminous components of Carina.  Interestingly, observational evidence
points toward the presence of unbound stars beyond the formal tidal
radius of this dwarf \citep{majewski05,munoz06b}, \citep[however see
discussion in ref.][]{penarrubia09}.

  These conclusions are subject to an important caveat. Our models
  do not take into account the gas content of the satellite, and
  therefore, all star formation activity that might occur after the
  infall time (our fiducial $t \sim 6$ Gyr) has been neglected. This
  could be of particular relevance for Carina, given the more extended
  star formation activity suggested by observations (e.g. Rizzi et
  al. 2003) \citep{rizzi03}. Simulations have shown that during
  pericenter passages episodes of star formation may be triggered for
  gas rich systems (Mayer et al.  2001,2006)
  \citep{mayer01a,mayer06}. Such events would re-establish a spatial
  (and presumably kinematical) bias between the old and newly formed
  stars.  This could certainly lower the efficiency of tides to
  spatially and kinematically mix the dwarf's stellar populations
  discussed in our models.  This is, however, unlikely to render
  invalid the results explored in Section 3, and their applicability
  to Carina. Notice that according to the recent models of Carina's
  star formation history, the bulk of stars is older than $\sim 5$ Gyr
  \citep{hurley-keller98, rizzi03}, which means that subsequent star
  formation episodes are unlikely to contribute to a large fraction of
  the total mass. Moreover, after the satellite has lost most of its
  dark matter mass, the mixing proceeds fast as the luminous
  components start to become stripped. Therefore, depending on the
  orbit it is still possible that mixing could have worked, at least
  partially, in Carina even if modest integration times, comparable or
  lower than the age of the dominant population, are invoked.

All these arguments are based on a model of a dwarf galaxy with two
initially segregated stellar components. However, neither the origin
nor how often galaxies are expected to show such configurations
has yet been properly understood.  Further constraints on the
validity (or not) of the extrapolations of our results to cases such
as that of Carina's will come from a better understanding of the
mechanisms able to generate metallicity gradients and spatial plus
kinematical segregation of stellar populations in the dwarf spheroidal
galaxies.

\section{Discussion and Conclusions}
\label{sec:concl}

We have studied, by means of N-body numerical simulations, the effects
of the tidal forces induced by a host galactic potential on a
multi-component satellite galaxy. The model satellite is set up {\it
ad$-$hoc} to initially have two spherical different stellar
populations, that are kinematically and spatially segregated.  A more
centrally concentrated and with lower velocity dispersion $C_1$
(reminiscent of a ``metal rich'' population) and a more extended, higher
velocity (``metal poor'') $C_2$ population. These two components are
deeply embedded within a dark matter halo that largely dominates the
total mass of the system. This satellite model roughly matches the
structural properties of classical Local Group dwarfs. We have
followed the evolution of such object in time for $t\sim 6$ Gyr on four
different orbits around a Milky Way-like host.

We find that the ability to distinguish kinematically the two stellar
components (set initially in our model to differ by $\sigma_2-\sigma_1
\sim 4$ km/s) is strongly dependent on the amplitude of the tides experienced by the
satellite during its orbit.  In cases where a significant amount of mass has been
removed, the velocity gap between the more concentrated (colder)
stellar population and the more extended (hotter) component can
decrease between $\sim$30-70\% of its initial value, thus partly erasing
the initial kinematical segregation between the stellar
populations. The magnitude of this effect is tightly related to the
tidal stripping experienced by the satellite, and in particular, the
removal of luminous mass is necessary for this effect to be
significant.  Such conditions are more easily obtained for orbits
restricted to the inner regions of the host potential, whose close
pericenter passages and short orbital periods promote the tidal
stripping of the satellite's particles.

We apply these ideas to Sculptor and Carina, two of the classical
dwarf galaxies of the Milky Way. Sculptor shows a stronger metallicity
gradient, together with a lower velocity dispersion for the more
centrally concentrated, metal rich stars compared to the metal poor
population. On the other hand, in Carina trends are considerably more
subtle, with no kinematical distinction between populations detected
to date. We argue that differences in the orbital paths of these two
dwarfs can be partially responsible for the weakening of the velocity
gap in Carina, while leaving Sculptor unaffected due to its more
external and longer period orbit.

Further validation of the tidal ``mixing'' scenario that has been
proposed in this {\it Paper} depends upon efforts from theoretical as
well as observational studies.  Comprehensive models to understand the
formation of the metallicity gradients as well as their likelihood in
dwarf galaxies are of fundamental relevance. The structure of the dark
matter halos of dwarf spheroidals (core vs cusped) might also play an
important role.  Cored profiles could provide less gravitational
support to oppose the tidal stripping forces compared to our cusped
models, perhaps changing (albeit we expect only quantitatively) the
timescales and relevance of the effects studied here.  Larger samples
of line of sight velocities for stars associated to dwarfs spheroidals
of the Local Group may improve the detectability of double stellar
components kinematically segregated in cases where such feature has
not yet been identified. Finally, our models predict that some degree
of mass depletion on the stellar components must take place in order
to considerably affect the kinematical segregation of dwarfs with
composite stellar populations. Deep photometric surveys mapping the
outskirts of these satellites, specially beyond their tidal radii,
should be able to provide definite clues on the existence (or not) of
tidally unbound stars associated to each of these objects, a necessary
(although not sufficient) condition for our models to apply.

\section*{Acknowledgements}

LVS would like to thank Eline Tolstoy for enlightening discussions.
LVS and AH gratefully acknowledge NWO and NOVA and the Coimbra Group
for financial support. We also thank both anonymous referees for useful
suggestions and comments.

\bibliography{references.bib}
\bibliographystyle{ieeetr}

\end{document}